\documentclass[10pt]{article}
\usepackage[english]{babel}
\usepackage{sansmathfonts}
\usepackage[T1]{fontenc}
\usepackage{amsmath,amsfonts,amssymb,amstext,amsmath,amsthm,amscd}
\usepackage{mathtools}
\usepackage[a4paper,top=2cm,bottom=2.5cm,left=2.5cm,right=2.5cm]{geometry}
\usepackage{breakcites}
\usepackage{mathtools}
\usepackage{enumitem}
\usepackage{booktabs}
\usepackage{mathtools}
\usepackage{mathrsfs}
\usepackage{dsfont}
\usepackage{enumitem}
\usepackage{graphicx}
\usepackage{subcaption}

\usepackage{wrapfig}
\usepackage{indentfirst}
\usepackage{latexsym}
\usepackage{sidecap}
\usepackage{booktabs}
\usepackage{hyperref}

  \usepackage{lipsum}
  \usepackage{cleveref}
  \usepackage[dvipsnames]{xcolor}
  \usepackage[nodisplayskipstretch]{setspace} 

\usepackage{tikz,tikz-3dplot}
\usepackage{ifthen}

\usetikzlibrary{patterns}%

\newcommand\myshade{85}
\colorlet{mylinkcolor}{violet}
\colorlet{mycitecolor}{YellowOrange}
\colorlet{myurlcolor}{RoyalBlue}

\hypersetup{
    colorlinks=true,
    linkcolor=myurlcolor,
    citecolor=mycitecolor!\myshade!black,
    filecolor=magenta,      
    urlcolor=magenta,
}
\usepackage{setspace}

\usepackage{enumitem}
\usepackage{mathrsfs}
\usepackage{textcomp}
\usepackage{braket}
\usepackage{colortbl}
\usepackage{bbm}
\usepackage{caption}
\usepackage{subcaption}
\usepackage{graphicx}
\usepackage{adjustbox}  
\usepackage{multirow}  
\usepackage{array}

\newcommand{\gsm}{GSM$^2\,$}
\newcommand{\mtgsm}{MT-GSM$^2\,$}

\begin{document}

\title{Integrating nano- and micrometer-scale energy deposition models for mechanistic prediction of radiation-induced DNA damage and cell survival}

\author{Giulio Bordieri$^{a,b}$ \and Marta Missiaggia$^{b,c}$ \and Gianluca Lattanzi$^{a,b}$ \and Carmen Villagrasa$^{d}$ \and Yann Perrot$^{d}$ \and Francesco G. Cordoni$^{e,b}$}
\date{}
\maketitle

\renewcommand{\thefootnote}{\fnsymbol{footnote}}
\footnotetext{{\scriptsize $^{a}$Department of Physics, University of Trento, via Sommarive 14, Trento, 38123, Italy}}
\footnotetext{{\scriptsize $^{b}$Trento Institute for Fundamental Physics and Application (TIFPA), via Sommarive 15, Trento, 38123, Italy}}
\footnotetext{{\scriptsize $^{c}$Department of Radiation Oncology, University of Miami Miller School of Medicine, Miami, 33136, FL, USA}}
\footnotetext{{\scriptsize $^{d}$Autorité de sûreté nucléaire et radioprotection (ASNR), Fontenay-aux-Roses, F-92262, France}}
\footnotetext{{\scriptsize $^{e}$Department of Civil, Environmental and Mechanical Engineering, University of Trento, via Mesiano 77, Trento, 38122, Italy}}

\footnotetext{{\scriptsize E-mail addresses: giulio.bordieri@unitn.it (Giulio Bordieri), marta.missiaggia@miami.edu (Marta Missiaggia), gianluca.lattanzi@unitn.it (Gianluca Lattanzi), carmen.villagrasa@irsn.fr (Carmen Villagrasa), yann.perrot@asnr.fr (Yann Perrot), francesco.cordoni@unitn.it (Francesco G. Cordoni)}}

\begin{abstract}
We present an integrated modeling framework that combines the Generalized Stochastic Microdosimetric Model (\gsm), used to predict cell survival fractions, with MINAS-TIRITH, a fast and efficient Geant4 DNA-based tool for simulating radiation-induced DNA damage in cell populations. This approach enables the generation of spatially and structurally resolved double-strand break (DSB) distributions, capturing key features such as damage complexity and chromosome specificity.
A novel application of the DBSCAN clustering algorithm is introduced to group DSBs at the micrometer scale. This allows the identification of physical aggregates of DNA damage and their association with subnuclear domains, providing a direct link to the cell survival probability as predicted by \gsm.
\\
\\
The model was validated using experimental data from HUVEC cells irradiated with 220 kV X-rays and H460 cells exposed to protons over a wide linear energy transfer (LET) range, from approximately $4~keV/\mu m$ to over $20~keV/\mu m$. Results show excellent agreement between simulations and experimental survival probabilities, making this one of the first consistent multi-scale models to bridge nanodosimetric and microdosimetric representations of radiation with biological outcomes such as cell survival. 
\\
\\
By incorporating the inherent stochastic nature of radiation-matter interactions, this framework effectively connects the physical properties of the radiation field to the biological response at the cellular level. Its accuracy across various radiation types and energies supports its potential for use in biologically optimized radiotherapy.\\
\textbf{Keywords:} Radiotherapy, Microdosimetry, Nanodosimetry, DNA damage, Biological modeling
\end{abstract}

\tableofcontents
\section{Introduction}
\label{0_Introduction}
\counterwithin{figure}{section}
\setcounter{figure}{0}

Approximately 50\% of patients diagnosed with localized malignant tumors undergo ionizing radiation therapy as part of their treatment regime~\cite{baskar2012cancer}, \cite{durante2019charged}. In recent decades, accelerated particles, mainly protons and carbon ions, have gained clinical interest as an alternative to conventional X-rays~\cite{ipe2010ptcog}. By the end of 2023, over $410,000$ patients had been treated with particle therapy worldwide, including nearly $350,000$ with protons (PTCOG 2024), making proton therapy the most widely adopted form of heavy ion treatment.
\\
\\
As particle therapy becomes increasingly prevalent, a deeper understanding of radiation's biological effects is essential for improving treatment outcomes and minimizing side effects. Accurate radiobiological models must capture the intrinsic stochasticity of radiation-matter interactions and the downstream biological response mechanisms. Such models are important in shaping clinical decisions across a spectrum of radiation qualities and delivery methods. Mechanistic mathematical models \cite{hawkins1994statistical, kase2006, vassiliev2012formulation, friedrich2013local, manganaro2017monte, vassiliev2017microdosimetry, inaniwa2018adaptation, cordoni2021generalized, cordoni2022multiple, mcmahon2017general} have emerged as powerful tools for describing radiation-induced biological damage. Unlike empirical or phenomenological approaches, \cite{mcnamara2015phenomenological, cordoni2023artificial, tian2022ion, flint2024empirical}, these models attempt to causally relate physical energy deposition patterns to cellular outcomes, offering improved generalizability and interpretability. While many traditional models focus on predicting treatment efficacy through cell survival curves, recent advances emphasize DNA double-strand break (DSB) induction and repair as fundamental determinants of cell fate~\cite{hu2025correlation, sakata2024prediction, hoang2024dsbandrepair}. Nevertheless, models of DSB induction and cell survival have traditionally evolved in parallel, with limited integration.
\\
\\
This study introduces an integrated multiscale framework that connects nanometric energy deposition and DSB induction with biological response and survival outcomes. Our approach extends the Generalized Stochastic Microdosimetric Model (\gsm)~\cite{cordoni2021generalized, cordoni2022cell, cordoni2023emergence, cordoni2024spatial, missiaggia2024cell, bordieri2024validation} by coupling it with MINAS-TIRITH, a Geant4-DNA-based simulator of radiation-induced DNA damage~\cite{thibaut2023minas, thibaut2023experimental}. The resulting \mtgsm framework enables spatially resolved simulation of DNA damage and repair across nanometer to micrometer scales. A key challenge in linking DNA damage modeling with cell survival prediction is reconciling their fundamentally different spatial and stochastic descriptions. DNA damage models typically operate at the nanometer scale, thus capturing local interactions around the DNA helix and accounting for stochastic intra-nuclear effects. In contrast, cell survival models often use micrometer-scale descriptions that aggregate energy deposition over entire nuclei or chromosomes, reflecting intra- and intercellular variability~\cite{friedrich2018dna}. Bridging these scales within a single computational model is essential to advance the predictive power of radiobiological models in clinical contexts.
\\
\\
The framework aims to characterize cellular responses across a spectrum of radiation qualities, including photon irradiation and clinically relevant proton energies. Rather than relying on the commonly used physical descriptor of proton radiation, i.e., Linear Energy Transfer (LET), which quantifies the energy deposited per unit path length, we adopt an alternative approach based on microdosimetry. Microdosimetry is a branch of physics that describes energy deposition at the micrometer scale, corresponding to the dimensions of DNA and chromosomes. Over the years, it has been shown to correlate strongly with various radiobiological endpoints by accounting for the stochastic nature of energy deposition within and between cells. This makes it a compelling alternative to LET for predicting radiobiological damage \cite{grun2019dose}. In clinical practice, a fixed Relative Biological Effectiveness (RBE) of 1.1 is often applied to proton therapy, implying that a given physical dose of protons is considered biologically equivalent to a 10$\%$ higher dose of photons \cite{paganetti2024nrg}. However, such simplifications fail to capture the complexity of radiation–matter interactions. For instance, it is well established that high-LET proton radiation induces more complex and clustered DNA damage, which is more difficult to repair and thus biologically more effective. Our approach replaces this oversimplification with a mechanistic model that directly links the characteristics of DNA damage to biological outcomes. This paves the way for potential integration into future treatment planning systems, offering a more accurate and biologically informed basis for optimizing radiation therapy.
\\
\\
Over the past decade, microdosimetry-based models, most notably the Microdosimetric Kinetic Model (MKM) and its generalizations \cite{bellinzona2021linking}, have demonstrated strong predictive accuracy for RBE. The MKM, in particular, is currently implemented in clinical practice for carbon ion therapy. A known limitation of many microdosimetric models, including \gsm, is using artificial “domains” to represent damage clustering within the cell nucleus. These domains, which influence the calculation of cell survival, are not biologically grounded and often calibrated empirically. For instance, \gsm currently uses a discrete domain structure, where DSB induction is estimated from track-structure simulations \cite{kundrat2020analytical}. While effective, this approach oversimplifies subnuclear heterogeneity and omits sub-micron scale spatial features that are biologically relevant. In particular, energy deposition events occurring below approximately 0.5 $\mu$m are typically ignored due to the lack of spatial resolution, despite their biological relevance. Furthermore, the size of these domains is often calibrated empirically against cell survival data, introducing a non-observable parameter not corresponding to any real biological structure. This limitation is not unique to \gsm but is inherent to all versions of the MKM, which remains one of the most widely used radiobiological models in clinical carbon ion therapy \cite{bellinzona2021linking}. To overcome these limitations, we integrate detailed DSB spatial data from MINAS-TIRITH \cite{thibaut2023minas, thibaut2023experimental}, a fast and biologically detailed Geant4-DNA-based simulator of radiation-induced DNA damage at the cell population level, directly into the \gsm framework. MINAS-TIRITH enables mechanistic modeling of energy deposition and DSB formation at nanometer resolution, capturing DNA damage's spatial and structural complexity with high fidelity. This detailed damage information is then processed using a novel DBSCAN-based clustering algorithm. This allows us to define biologically meaningful subnuclear domains from first principles and enables accurate modeling of LET-dependent effects. For instance, our model captures RBE variations directly from physical and biological mechanisms, without empirical correction factors.
\\
\\
The final model integrates MINAS-TIRITH with \gsm (\mtgsm) and uniquely combines the accurate and rapid computation of radiation-induced DNA damage distributions at the nanometer scale with precise DNA repair and inter-cell stochasticity as described by microdosimetry. MINAS-TIRITH simulates the DSB distribution, including the nuclear coordinates, the complexity of the damage, and the specific chromosome on which it occurs. We implement a novel approach that groups DSBs into $\mu m$-size clusters using DBSCAN, linking physical data to biologically meaningful nuclear subdomains for the first time.
\\
\\
The new model was first validated against endothelial cell (HUVEC) irradiation experiments with a Small Animal Radiation Research Platform (SARRP) using 220~kV X-rays \cite{paget2019multiparametric}. It was then further tested on protons using H460 cells across a wide LET range. We consider the experiments performed in \cite{patel2017optimization} and \cite{bronk2020mapping} with protons and reproduce their experimental setup in Geant4 \cite{agostinelli2003geant4,allison2006geant4,allison2016recent}. These datasets were selected due to their high measurement accuracy and comprehensive documentation of physical parameters, especially microdosimetric quantities such as the dose-mean lineal energy. Notably, results from proton irradiation in \cite{bronk2020mapping}, show an interesting effect in the high-LET regime, with RBE values exceeding the standard reference of $1.1$ by a factor around three.
\\
\\
By linking nanodosimetric energy deposition patterns with cellular-scale outcomes, this study provides one of the first multiscale, mechanistically grounded models of radiation-induced cell inactivation. The \mtgsm framework offers computational efficiency, biological interpretability, and broad applicability for the modeling and optimization of advanced radiotherapy strategies.

\section{Materials and methods}
\label{1_Materials_Methods}

\subsection{MINAS-TIRITH Computational Toolkit}
\label{Subsubsection: MINAS-TIRITH}
The MINAS-TIRITH computational toolkit \cite{thibaut2023minas, thibaut2023experimental} was developed to exploit the power of accurate Track Structure Monte Carlo simulations to characterize the nanodosimetric distribution of DNA damage induced by ionizing radiation within a cell population. This approach captures the stochastic nature of radiation-matter interactions by incorporating both microdosimetric and nanodosimetric features. To achieve computational efficiency, MINAS-TIRITH utilizes precomputed databases of microdosimetric parameters -namely energy deposited per track length distributions- and DNA damage distributions generated with the Geant4-DNA Monte Carlo Toolkit \cite{incerti2010geant4,incerti2010comparison,incerti2018geant4,tran2024review}. These databases are currently available for electrons $[1~keV - 1 MeV]$, protons $[10~keV - 21 MeV]$, and helium ions $[10~keV - 21 MeV]$.
\\
\\
MINAS-TIRITH estimates the DNA damage topology within a cell population subjected to a given irradiation scenario, achieving significantly reduced computational times relative to full track-structure simulations. The default nucleus model is an elliptical cylinder with a height of $2~\mu m$, a half-major axis of $9.5~\mu m$ and a half-minor axis of $5.1~\mu m$, representing average cell nuclear dimensions.
\\
\\
To simulate the radiation-induced damage, MINAS-TIRITH requires the following input parameters:

\begin{enumerate}[label=(\roman*)]
    \item Number of cells in the population ($N_{\text{cells}}$)
    \item Macroscopic absorbed dose delivered to the cell population ($D_{\text{abs}}$)
    \item Energy spectra of incident particles
    \item Angular distributions of incident particles
    \item Relative particle weights (fractional abundance).
\end{enumerate}
Before running MINAS TIRITH, a phase space at the interface between the cell nuclei and the culture medium must be computed. This phase space describes the particle types, energies, and angles at the boundary and is obtained through Geant4 Monte Carlo simulations that accurately replicate the experimental irradiation setup. 
Given the macroscopic absorbed dose $D_{\text{abs}}$, MINAS-TIRITH assigns a stochastic specific energy $z_n$ to each individual cell nucleus in accordance with microdosimetric principles. The number and topology of DNA damage events within each cell are then sampled from precomputed distributions based on  $z_n$. 
The simulation outputs, generated for each cell, list all ionizing radiation events within the nucleus. For each DSB-type damage, the following attributes are recorded:
\begin{enumerate}[label=(\roman*)]
    \item Cartesian coordinates ($X,Y,Z$) of the DNA damages within the cell nucleus;
    \item ID of the specific chromosome on which the damage is created;
    \item Complexity index $i \geq 0$ denoting the additional number of single-strand breaks associated with the same DSB;
    \item Origin of the DSB-type damage in terms of their composing strand breaks (SB) (physical, chemical, or hybrid).
\end{enumerate}
Here, physical SBs result from direct ionizations radiation on picosecond timescales, while chemical SBs originate from indirect interactions, primarily via hydroxyl radicals from water radiolysis. Hybrid DSB-type damages involve combination of SBs with contributions from both processes, reflecting the complex interplay of direct and indirect mechanisms.

\subsection{The Generalized Stochastic Microdosimetric Model (GSM$^2$)}
\label{Subsubsection: GSM2 model}
\gsm, first formulated in \cite{cordoni2021generalized}, is a stochastic radiobiological model designed to account for the randomness in DNA damage induction, repair, and fixation. The model partitions the cell nucleus into $N_d$ independent domains. DNA lesions are cathegorized into sublethal and lethal lesions, denoted respectively $X$ and $Y$. Sublethal lesions may be repaired, directly converted into lethal lesions, or interact with other sublethal lesions, producing a lethal one. On the other hand, lethal lesions are unrecoverable and lead to cell inactivation. The following reaction pathways describe the dynamic behavior of sublethal lesions:

\begin{equation}
    X \xrightarrow{r} \emptyset ~~~~~~~
    X \xrightarrow{a} Y         ~~~~~~~
    X + X \xrightarrow{b} Y
    \label{Eq.: GSM2 pathways},
\end{equation}

where $\emptyset$ represents the set of healthy cells, $r$ is the repair rate, $a$ is the spontaneous conversion rate to lethal lesions, and $b$ is the rate of pairwise interactions and recombination into a lethal lesion~\cite{cordoni2021generalized, cordoni2022cell}.
In this work, \gsm is integrated with MINAS-TIRITH outputs, bypassing the need for artificial spatial domains. The initial damage distribution is computed from simulated DSB coordinates without relying on domain discretization.

\subsection{Chromosome-aware DSB clustering using DBSCAN}
\label{Subsection: DBSCAN algorithm}
To bridge the gap between nanoscale lesion data and the \gsm model's spatial structure, we introduce a chromosome-aware DSB clustering algorithm based on the DBSCAN method~\cite{ester1996density,ester1997density,francis2011simulation}. Unlike conventional approaches relying on arbitrary geometric domains, our method identifies biologically plausible DSB clusters using proximity and chromosomal identity.
The clustering algorithm operates under the following rules:
\begin{enumerate}[label=(\roman*)]
    \item a cluster must contain at least one DSB;
    \item DSBs weighted by their complexity index $i$, such that indices $i$ and $i+1$ contribute equally to the total number of damages growing in the sequence $0, 1, 2, \ldots$;
    \item DSBs must lie within $1~\mu m$ of each other to be considered part of the same cluster;
    \item  DSBs must originate from the same chromosome to belong to the same cluster.
\end{enumerate}
The $1~\mu m$ proximity threshold aligns with microdosimetric conventions and the approximate dimensions of subnuclear structures. Importantly, this parameter is not optimized or fitted to data but is chosen a priori to remain consistent with established theoretical frameworks.

\subsection{Survival Probability Estimation}
\label{Subsection: survival calculation}
Once DSB clusters are defined, they serve as functional analogs to domains in \gsm. For each simulated nucleus irradiated with a macroscopic average dose $D_{\text{abs}}$ receiving a stochastic specific energy $z_n$, the survival probability is computed using the expression~\cite{cordoni2022cell}:

\begin{align}
    \label{Eq. Sn(Dabs)}
    S_n(z_n|D_{\text{abs}}) &= \prod_{j=1}^{N_d} \prod_{x=1}^{x_j} \dfrac{rx}{(a+r)x + bx(x-1)},
\end{align}

where $x_j$ denotes the number of sublethal lesions in the $j$-th cluster and $N_d$ is the number of DSB clusters determined via DBSCAN, as reported in Section \ref{Subsection: DBSCAN algorithm}. As previously mentioned, DSB-type damages with $i$ and $i+1$ account for the same number of sublethal lesions, starting with one sublethal lesion corresponding to a DSB-type damage with $i = 0$ or $i = 1$.
\\
\\
This calculation is repeated for all simulated cells ($N_{\text{cell}}$), and the population-level survival probability is obtained as:
\begin{align}
    \label{Eq. Sn(Dabs)}
    S(D_{\text{abs}}) &= \dfrac{1}{N_{\text{cell}}}\sum_{n=1}^{N_{\text{cell}}} S_n(z_n|D_{\text{abs}}).
\end{align}
The specific energy $z_n$ for each nucleus is sampled from the microdosimetric multi-event distribution produced by MINAS-TIRITH \cite{thibaut2023minas}.

\subsection{Biological endpoint: $RBE_{10}$ estimation}
\label{Subsection: RBE[10]}

To quantify the relative biological effectiveness (RBE), we compute the RBE at $10\%$ survival (RBE$_{10}$), defined as:
\[
\mbox{RBE}_{10}= \left . \frac{D_{\mbox{X-rays}}}{D_{\mbox{ions}}} \right |_{\mbox{SF} = 10\%},
\]

where $D_{\text{X-rays}}$ and $D_{\text{ions}}$ are the doses corresponding to a $10\%$ survival fraction under the different irradiation types.

To calibrate the \gsm biological parameters ($a$, $b$, $r$), we minimize the normalized mean square error (NMSE) between the predicted and the experimental cell survival, normalized by the experimental cell survival, as described in \cite{missiaggia2024cell, bordieri2024validation}:
\begin{equation}
    \label{Eq. NMSE}
    \text{NMSE} := \sum_D \dfrac{1}{S(D)} \left[S(D) - S_{\text{MT-GSM$^2$}}(D) \right]^2,
\end{equation}

where $S(D)$ denotes the experimental surviving fraction at a dose $D$, and $S_{\text{MT-\gsm}}(D)$ the corresponding model predictions. The minimization uses the Limited Memory Broyden-Fletcher-Goldfarb-Shanno algorithm with box constraints (L-BFGS-B) \cite{byrd1995limited}. Additionally, $a$, $b$, and $r$ are imposed to be positive, and the algorithm is run with starting points $a = 0.1$, $b = 0.1$, and $r = 2$. 
Notably, only the \gsm parameters are fitted, while the DBSCAN clustering parameters—distance threshold ($1,\mu$m) and minimum cluster size (1 DSB)—are fixed and not subject to optimization. This ensures that the MT-\gsm model retains a biophysically grounded and parameter-free definition of damage domains. This distance is chosen to align with all the other mechanistic models founded on microdosimetry \cite{hawkins2003microdosimetric,bordieri2024validation}.

\subsection{Radiobiological data and Geant4 Monte Carlo simulations}
\label{Subsection: Experimental works}
The model is validated against both X-ray and proton irradiation data. For X-rays, we use experimental results from the Autorité de Sûreté Nucléaire et de Radioprotection (ASNR) involving $220~kV$ X-ray exposures on survival curves obtained by clonogenic assay as described in~\cite{paget2019multiparametric}.
\\
\\
We refer to a radiobiological study conducted at the MD Anderson Cancer Center (University of Texas) with the H460 cell line for proton irradiation. This highly radiosensitive non-small cell lung carcinoma model is well-suited for probing the variability in cell survival due to differences in radiation quality. Increased radiosensitivity enhances the biological contrast between different LET conditions, making it ideal for assessing the impact of track structure. The reference experiment covered a broad LET range, from approximately $4~keV/\mu m$ to over $20~keV/\mu m$, with full experimental details available in \cite{patel2017optimization,bronk2020mapping}.
\\
\\
To replicate the irradiation conditions used in these experiments, we implemented the full experimental setup within the Geant4 MC simulation toolkit (version 11.2.0) \cite{agostinelli2003geant4}. This allowed us to accurately score the phase space of primary and secondary particles, including energy and angular distributions, entering the cell nuclei, which is required as input for the MINAS-TIRITH simulations. The electromagnetic interactions were modeled using the \textit{G4EmStandardPhysics option4} physics list, while hadronic interactions were described using the \textit{QGSP BIC HP} model. A production threshold of $0.1~keV$ and a spatial cutoff of $50~nm$ were used to control the creation of secondary particles. We simulated a minimum of $3 \times 10^5$ primary particles for each irradiation condition to generate statistically robust phase space data while maintaining reasonable computational efficiency. All available proton experiments were simulated, except those whose phase space results in an energy spectrum exceeding the current limits of the MINAS-TIRITH database.
\\
\\
To compute the RBE values for protons at different LETs, the reference SF curves were taken from photon irradiations using $^137$Cs (Cesium-137), following the experimental protocol described in \cite{guan2015spatial}. In table \ref{tab:GSM2_parameters} we show the linear-quadratic (LQ) model parameters used in RBE computation for H460 cells, ensuring consistency with the experimental datasets.
\\
\\
MINAS-TIRITH was then used to calculate the DSB distribution for each irradiation scenario. For X-rays, simulations were performed for macroscopic absorbed doses of  $[0.5,1,2,3,4,5]~Gy$. Five dose points were chosen for each LET value for protons, ensuring that the maximum dose did not exceed the highest dose used in the corresponding experimental dataset. All MINAS-TIRITH computations were performed for a population of  $N_{\text{cells}} = 2000$ simulated cell nuclei.

\section{Results}
\label{2_Results}
\counterwithin{figure}{section}
\setcounter{figure}{0}

The \mtgsm framework was validated against surviving fraction (SF) data for two types of radiation: 220 kV X-rays and protons across a wide LET range. X-ray irradiation was modeled for HUVEC endothelial cells, while proton irradiation experiments were simulated for H460 lung cancer cells.

\subsection{DSB clustering under varying dose and LET conditions}
\label{Subsubsection: DSB clustering}
\begin{figure}[!h]
    \centering
    \begin{subfigure}[b]{0.6\textwidth}
        \centering
        \includegraphics[width=\textwidth]{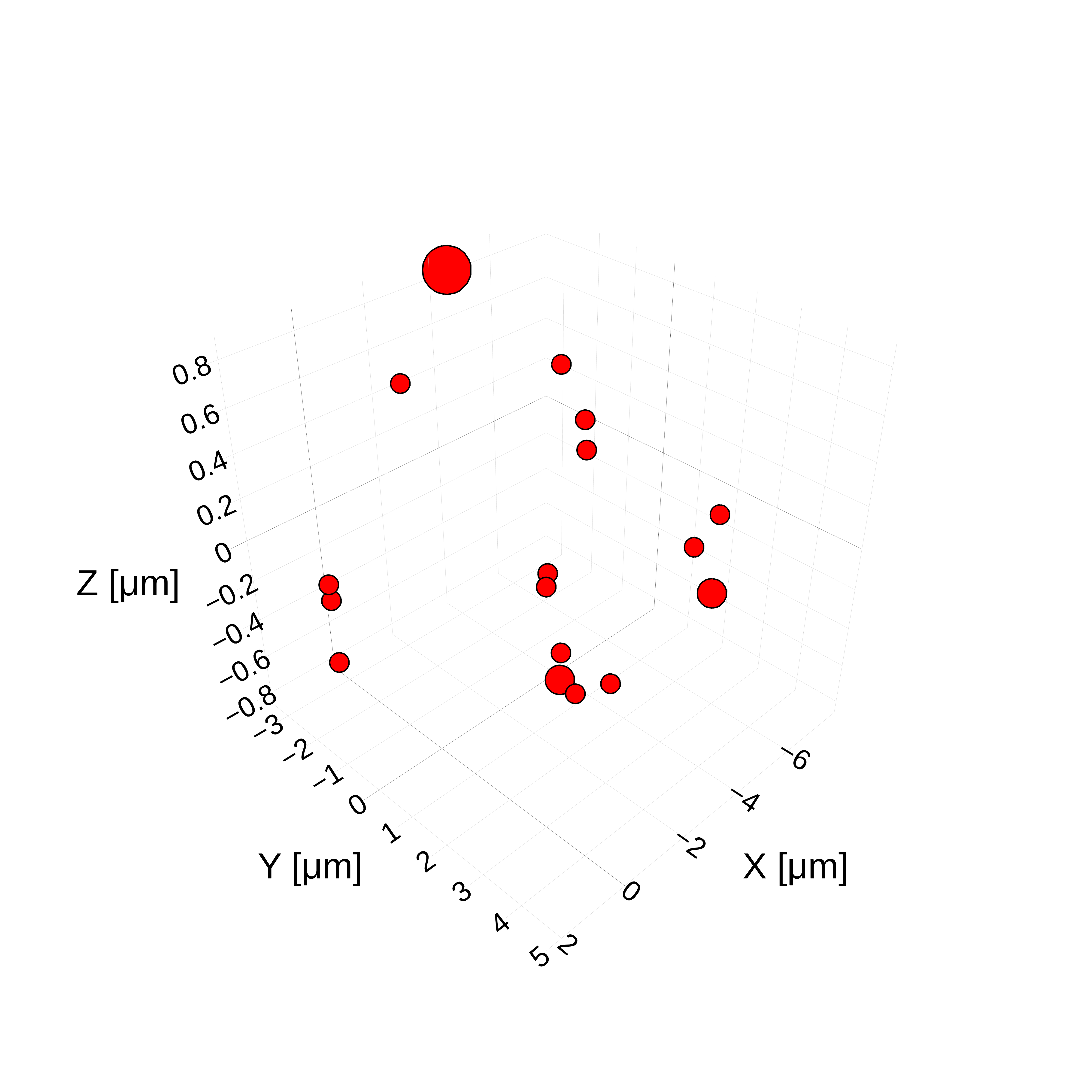}
        \caption{3D mapping of DSB generated by MINAS-TIRITH}
        \label{Fig. DSBmap_default}
    \end{subfigure}
    \hfill
    \begin{subfigure}[b]{0.6\textwidth}
        \centering
        \includegraphics[width=\textwidth]{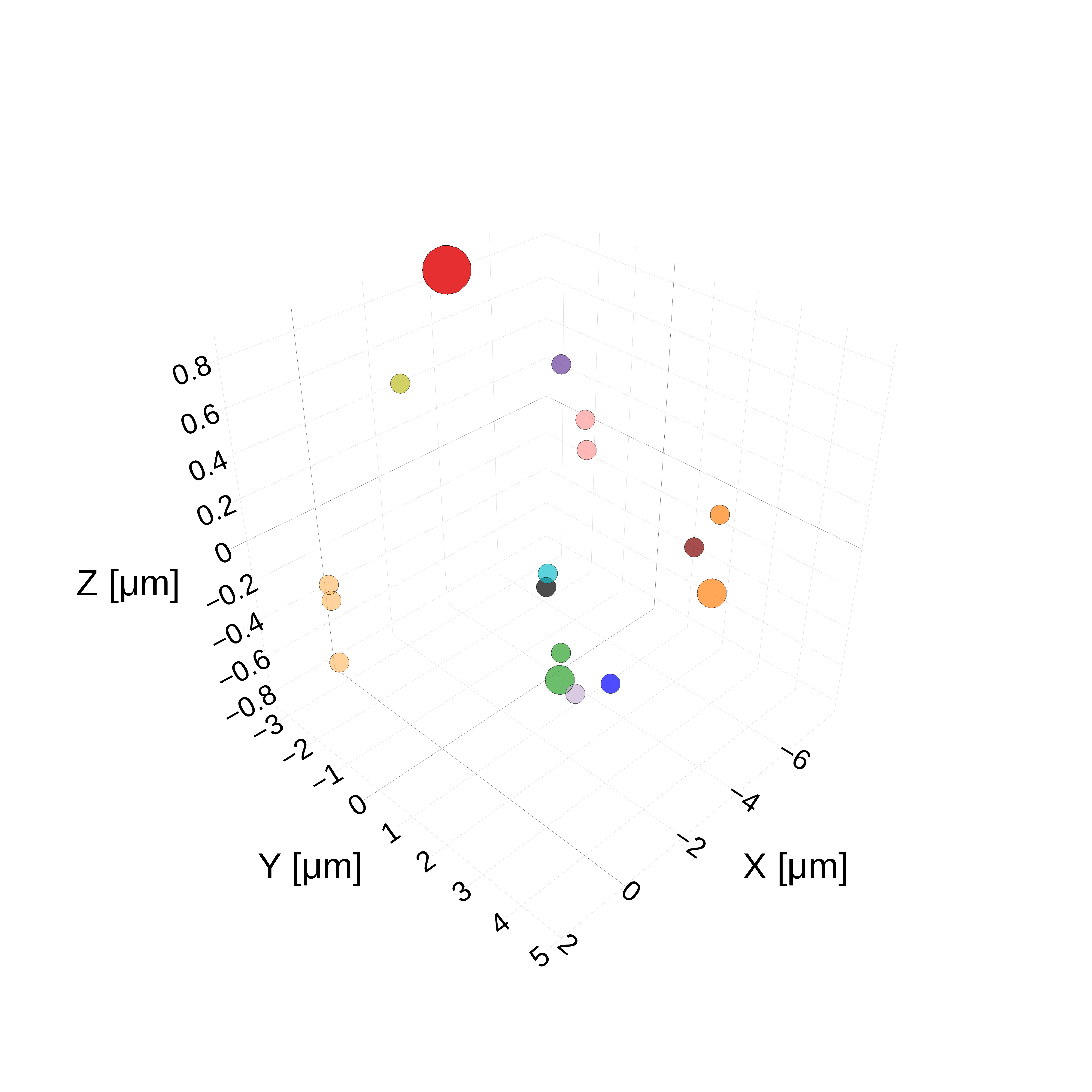}
        \caption{Clustering process described in subsection \ref{Subsection: DBSCAN algorithm}.}
        \label{Fig. DSBmap_DBSCAN}
    \end{subfigure}
    \caption{3D visualization of DSBs and clustering results. Each circle represents a DSB, with size reflecting complexity.}
    \label{Fig. DSBmaps_combined}
\end{figure}

Figures \ref{Fig. DSBmap_default} and \ref{Fig. DSBmap_DBSCAN} present an example of a simulated irradiation event within a single cell nucleus. Figure \ref{Fig. DSBmap_default} shows 3D mapping of DSB-type damages as red dots, where the size of each dot is proportional to its complexity index. Figure \ref{Fig. DSBmap_DBSCAN} illustrates the clustering process using the DBSCAN algorithm, with each cluster assigned to a unique color.

\begin{figure}[!h]
    \centering
    \begin{subfigure}[b]{0.48\textwidth}
        \centering
        \includegraphics[width=\textwidth]{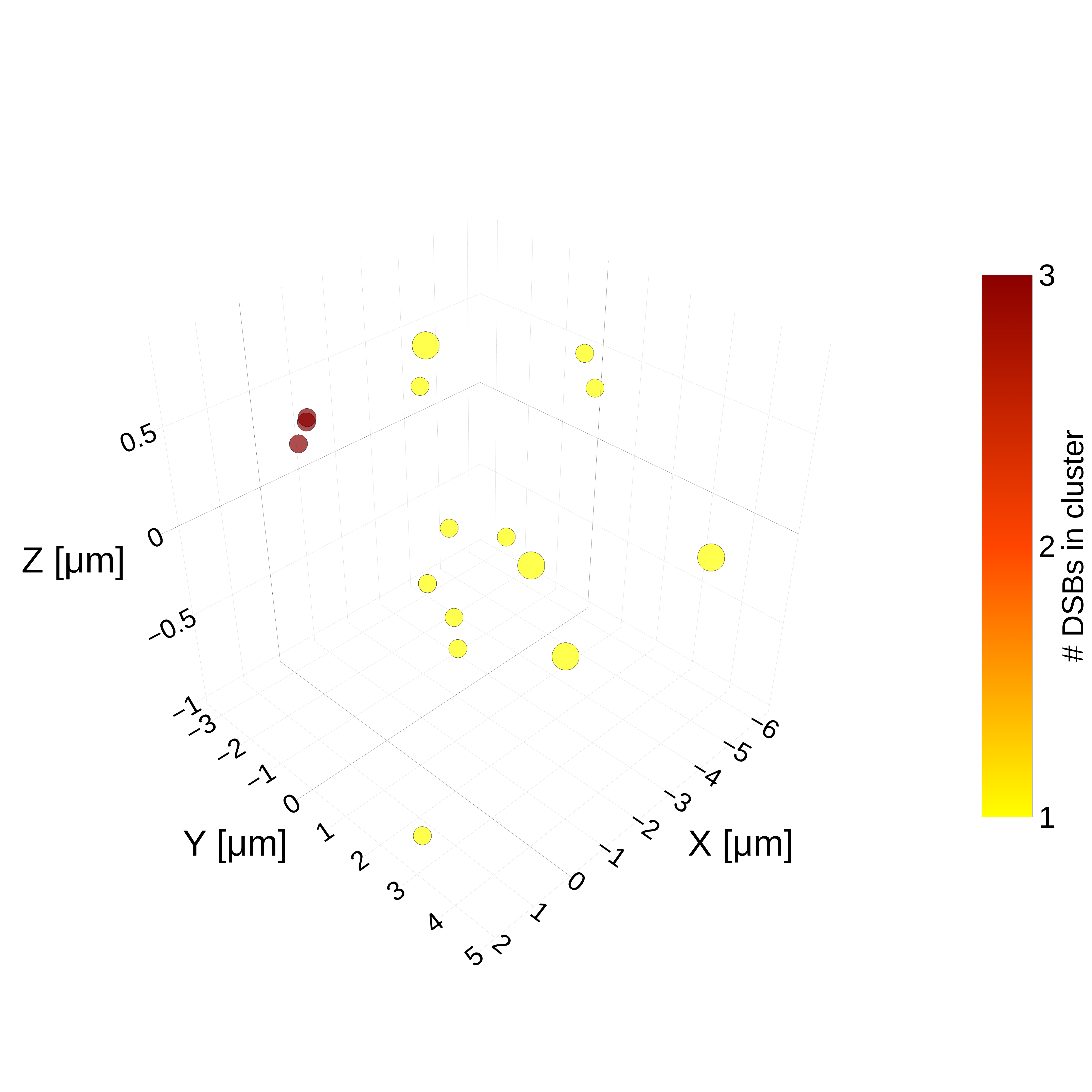}
        \caption{LET $= 5~keV/\mu m$, $D_{\text{abs}} = 1~Gy$}
        \label{Fig. DSBmap_lowLET_dose1}
    \end{subfigure}
    \hfill
    \begin{subfigure}[b]{0.48\textwidth}
        \centering
        \includegraphics[width=\textwidth]{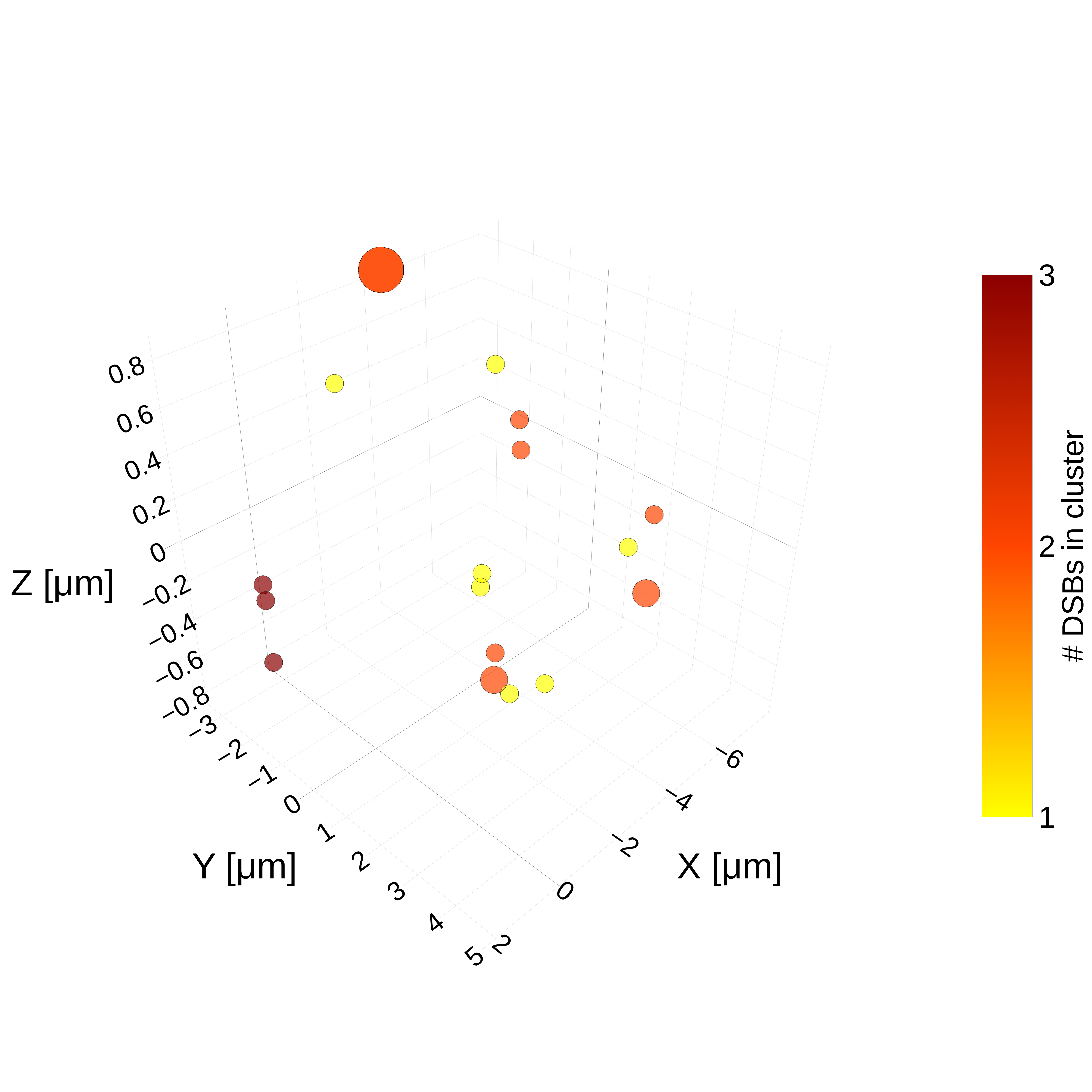}
        \caption{LET $= 20~keV/\mu m$, $D_{\text{abs}} = 1~Gy$}
        \label{Fig. DSBmap_highLET_dose1}
    \end{subfigure}
    \hfill
    \begin{subfigure}[b]{0.48\textwidth}
        \centering
        \includegraphics[width=\textwidth]{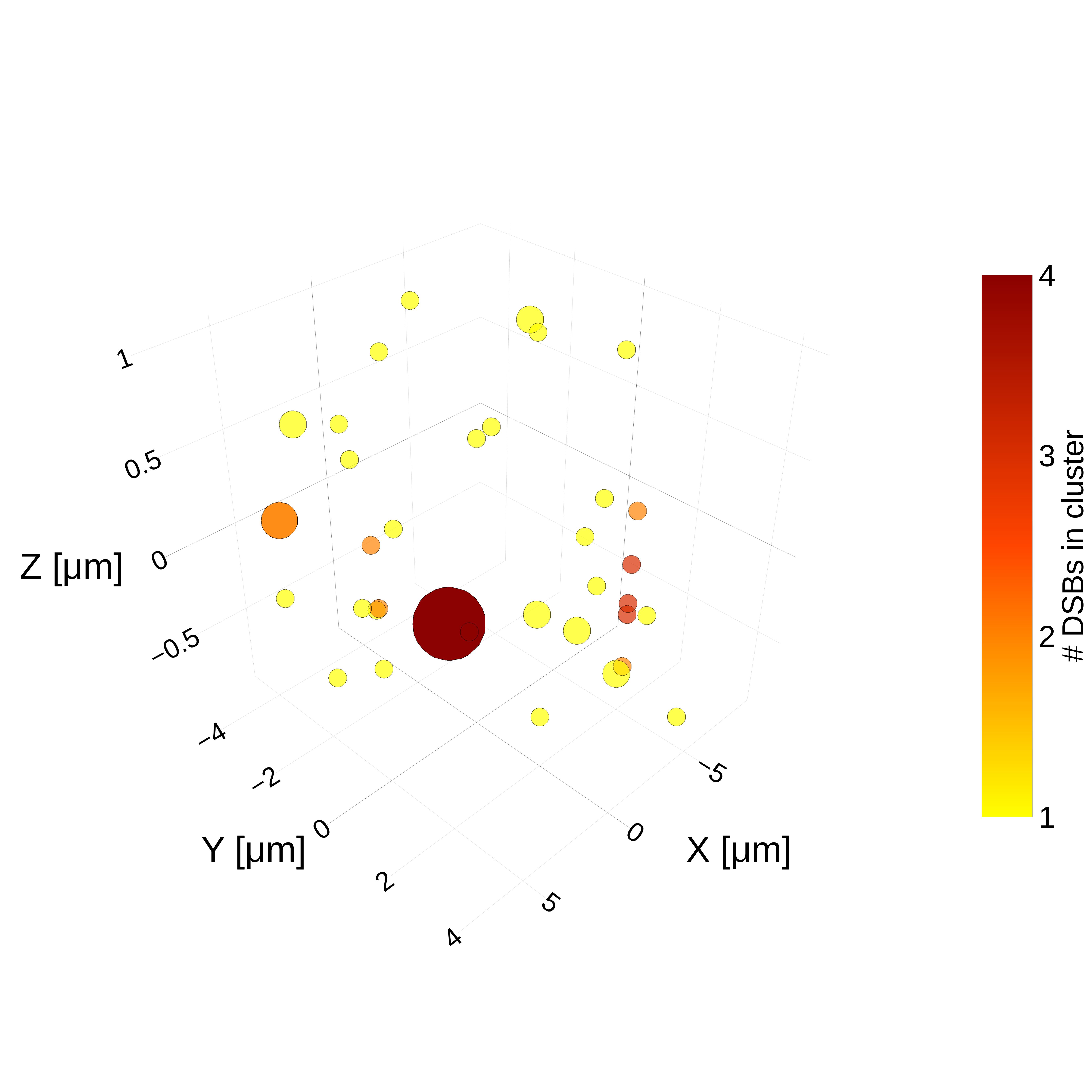}
        \caption{LET $= 5~keV/\mu m$, $D_{\text{abs}} = 4~Gy$}
        \label{Fig. DSBmap_lowLET_dose4}
    \end{subfigure}

    \caption{Examples of clustered DSB distribution under varying LET and dose conditions.}
    \label{Fig. DSBmaps}
\end{figure}

Further examples of DBSCAN clustering applied to MINAS-TIRITH outputs are shown in Figures \ref{Fig. DSBmap_lowLET_dose1}, \ref{Fig. DSBmap_highLET_dose1} and \ref{Fig. DSBmap_lowLET_dose4}. Three representative scenarios are depicted: (a) low LET and low dose (LET $= 5~keV/\mu m$ and $D_{\text{abs}} = 1~Gy$), (b) high LET and low dose (LET $= 20~keV/\mu m$ and $D_{\text{abs}} = 1~Gy$), and (c) low LET and high dose (LET $= 5~keV/\mu m$ and $D_{\text{abs}} = 4~Gy$). These configurations highlight how both radiation quality and dose influence the spatial distribution and complexity of DSBs.
Each circle represents a DSB-type damage with size proportional to its complexity index: clusters are colored to indicate their dimensionality.

\subsection{Survival prediction for 220 kV X-rays on HUVEC Cells}
\label{Subsubsection: X-rays results}

\begin{figure}[!h]
    \centering
    \begin{subfigure}[b]{0.8\textwidth}
        \centering
    \includegraphics[width=\textwidth]{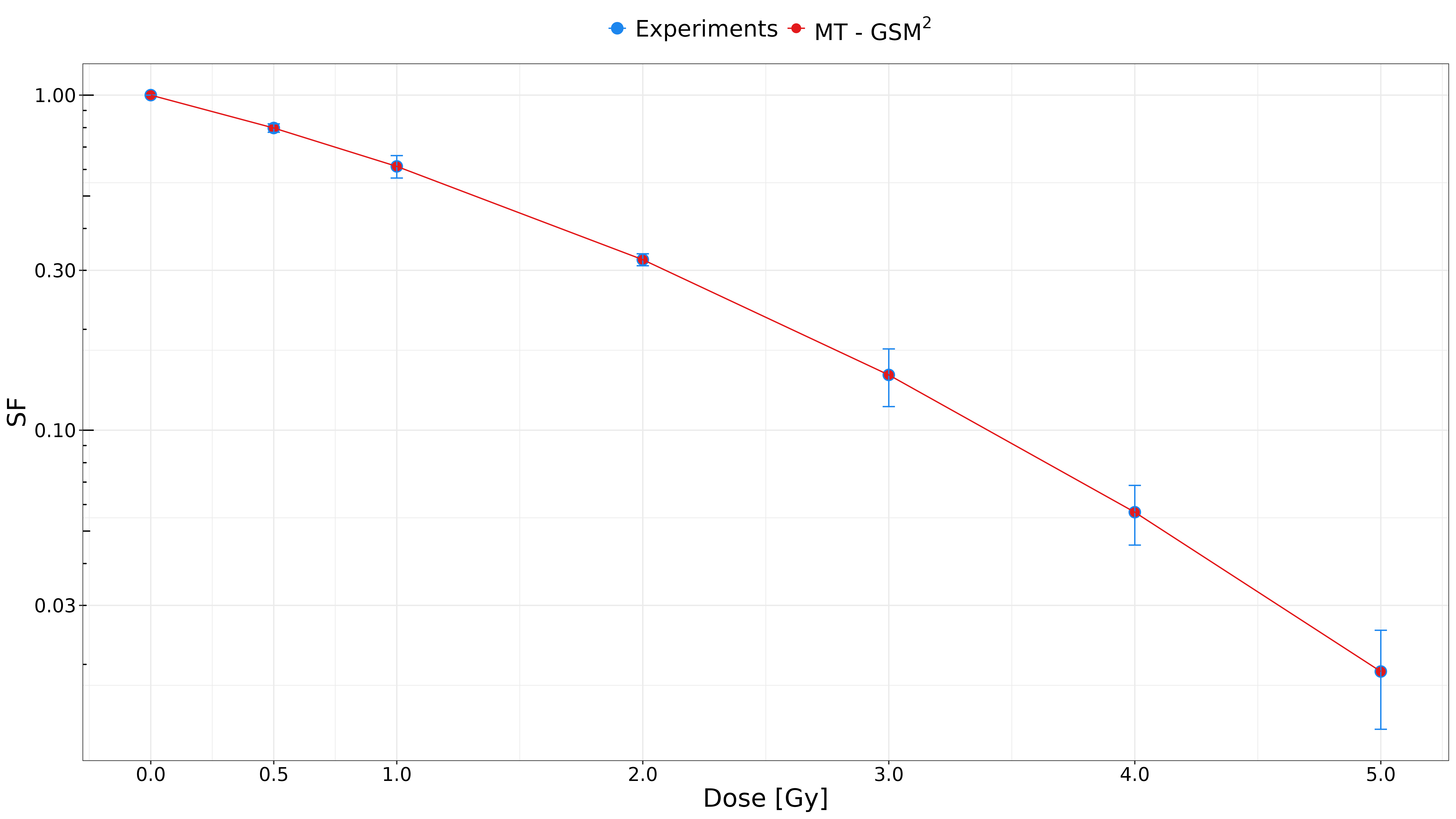}
    \caption{SF curve for HUVEC endothelial cells irradiated with 220 kV X-rays. Blue points show data from the experimental measurements performed at ASNR, while the predictions from MINAS-TIRTIH and \gsm are in red.}
    \label{Fig. SF_Xrays}
    \end{subfigure}
    \hfill
    \begin{subfigure}[b]{0.8\textwidth}
            \centering
    \includegraphics[width=\textwidth]{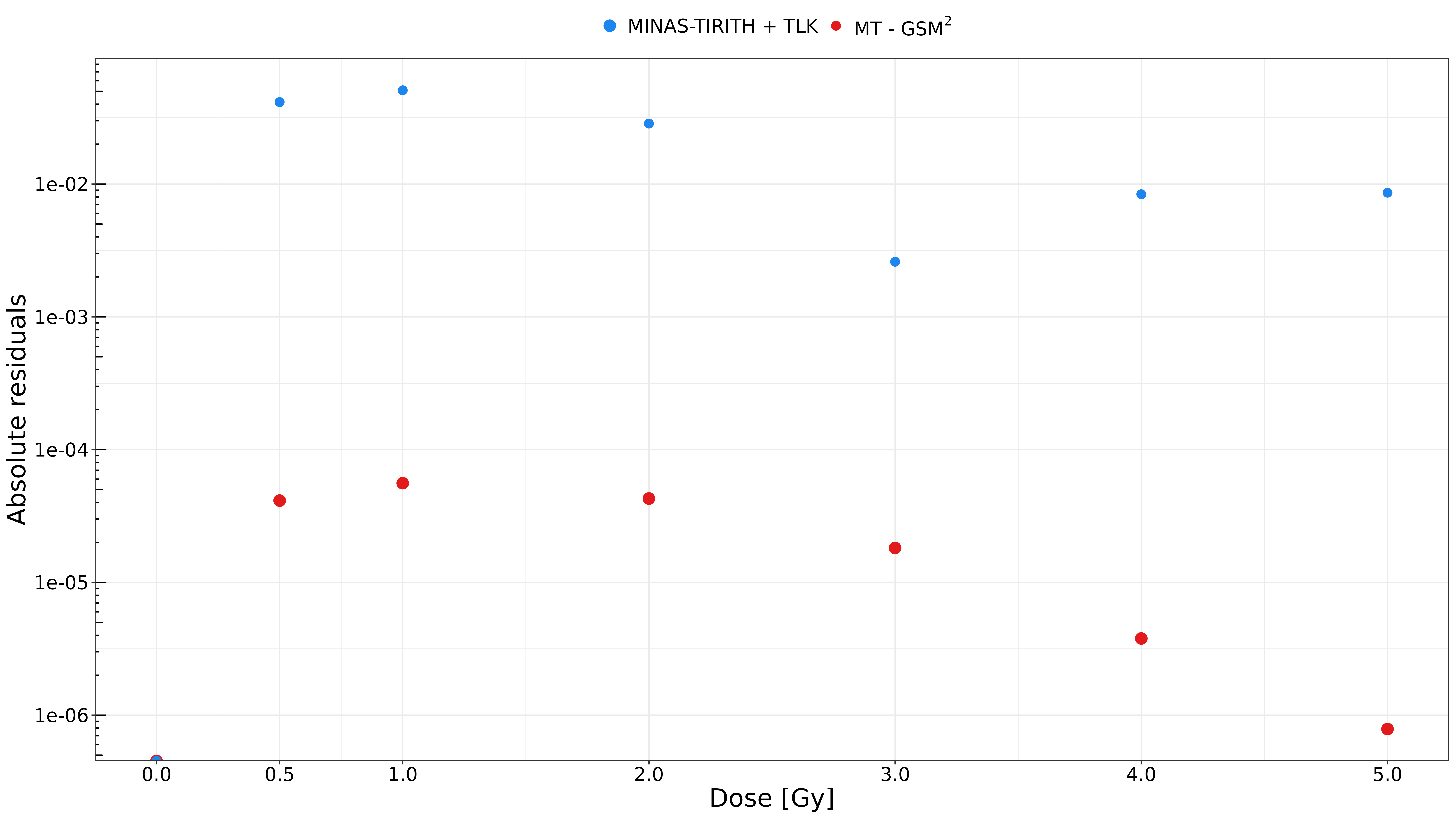}
    \caption{Absolute residuals from SF data for HUVEC endothelial cells irradiated with 220 kV X-rays. Blue points show the predictions from the Two Lesions Kinetic Model as currently implemented in Geant4-DNA. Red points show the performance from \mtgsm.}
    \label{Fig. Res_Xrays}
    \end{subfigure}
    \caption{MT-GSM$^2$ prediction for HUVEC endothelial cells irradiated with 220 kV X-rays.}
    \label{Fig. XRays}
\end{figure}

The model's predicted SF curve (\mtgsm, red) is compared to experimental data (blue) from Paget et al. \cite{paget2019multiparametric} in Figure \ref{Fig. SF_Xrays}.
The results demonstrate strong agreement across all tested doses. 
Residual analysis (Figure \ref{Fig. Res_Xrays}) shows that \mtgsm predictions (red) offer a substantially improved fit over the Two-Lesion Kinetic Model \cite{stewart2001two} (TLK, blue) currently implemented in Geant4-DNA, reducing absolute residuals by over two orders of magnitude.

\subsection{RBE$_{10}$ predictions for protons on H460 cells}
\label{Subsubsection: protons results}
\begin{figure}[!h]
    \centering
    \includegraphics[scale=0.5]{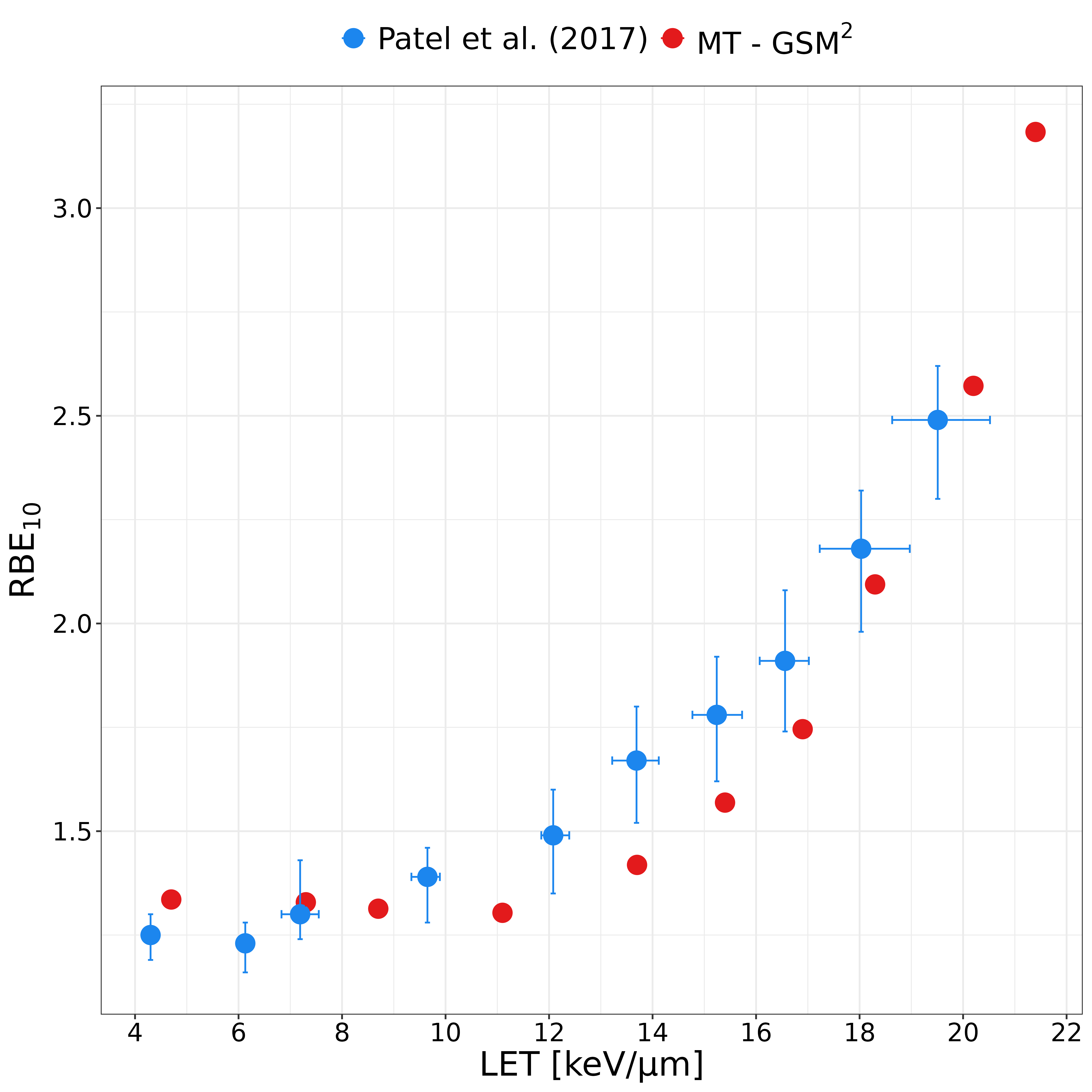}
    \caption{Comparison of RBE$_{10}$ values for protons irradiation on H460 cells. Red circles show the predicted values by the model, and the blue circles show experimental data from \cite{patel2017optimization}.}
    \label{Fig. RBE_H_H460}
\end{figure}

Figure \ref{Fig. RBE_H_H460} compares RBE$_{10}$ values predicted by \mtgsm (red) with experimental results from Patel et al. \cite{patel2017optimization} (blue) for H460 cells across a range of LET values. The model accurately captures the increase in biological effectiveness with LET.

\subsection{DSB classification from MINAS-TIRITH}
\label{Subsubsection: MINAS-TIRITH results}
\begin{figure}[!h]
    \centering
    \includegraphics[width=\textwidth]{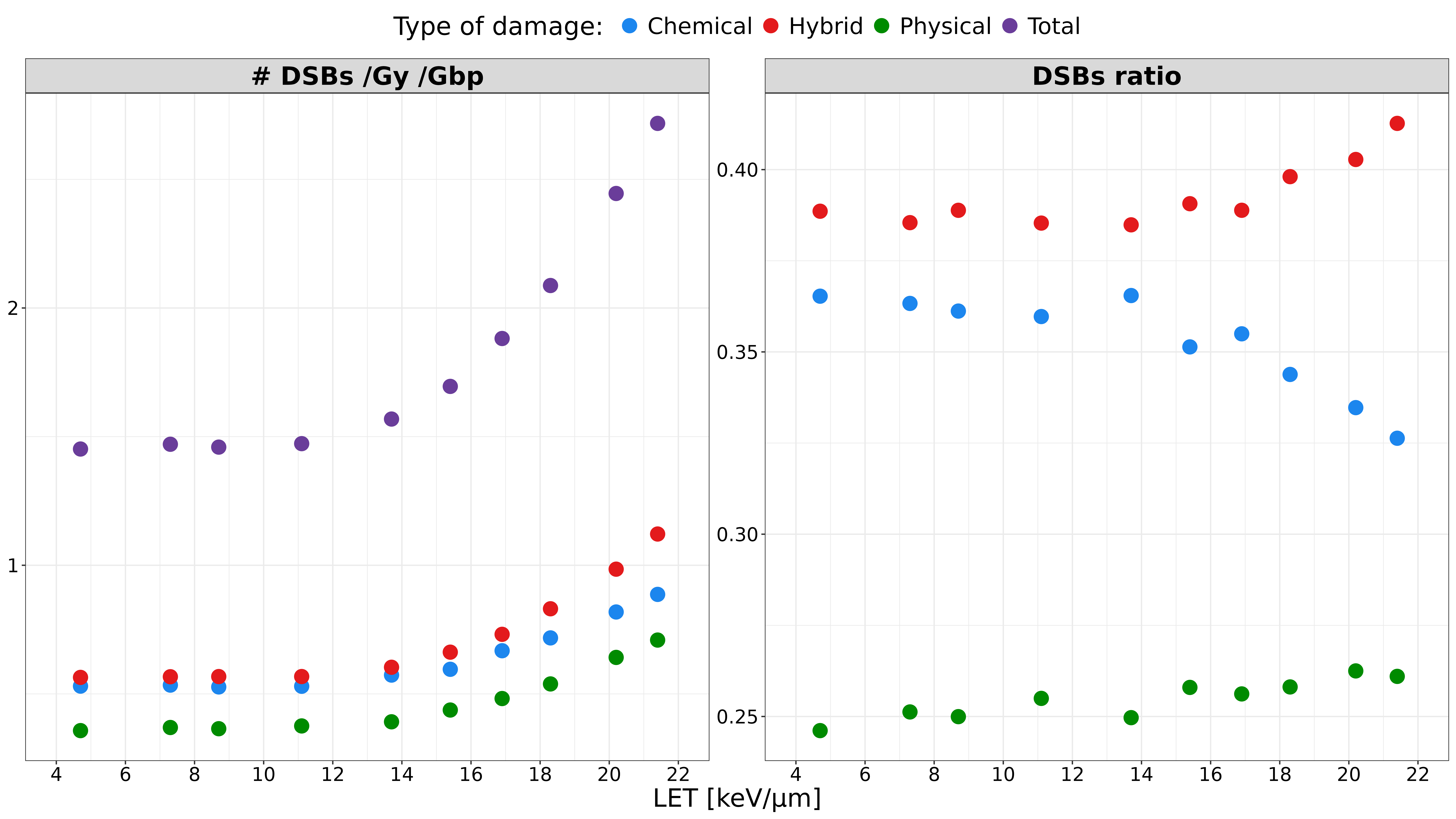}
    \caption{\textbf{Left panel}: number of DSBs per Gray per Giga base-pairs against LET from MINAS-TIRITH. DSBs are divided into types describing the origin of each damage. Purely physical DSBs are shown in green, purely chemical in blue, while damages caused by a hybrid process are represented in red. The purple dots show the total amount of damage. \textbf{Right panel}: relative proportion of DSB types versus LET. The color code is as specified for the left panel.}
    \label{Fig. DSB_dir_indir}
\end{figure}

The total number of DSBs per Gray per Giga base-pairs as a function of LET (purple), and the classification of their composing strand breaks by origin (physical in green, chemical in blue, or hybrid in red) is shown in Figure \ref{Fig. DSB_dir_indir}. The left panel shows absolute DSB yields, while the right panel presents the relative fraction of each DSB type. As LET increases, the contribution of chemically induced damage decreases, while the share of physical and hybrid DSBs increases.

\subsection{Impact of nanodosimetry on biological effectiveness}
\label{Subsubsection: nanodosimetry impact}
\begin{figure}[h]
    \centering
    \includegraphics[scale=0.5]{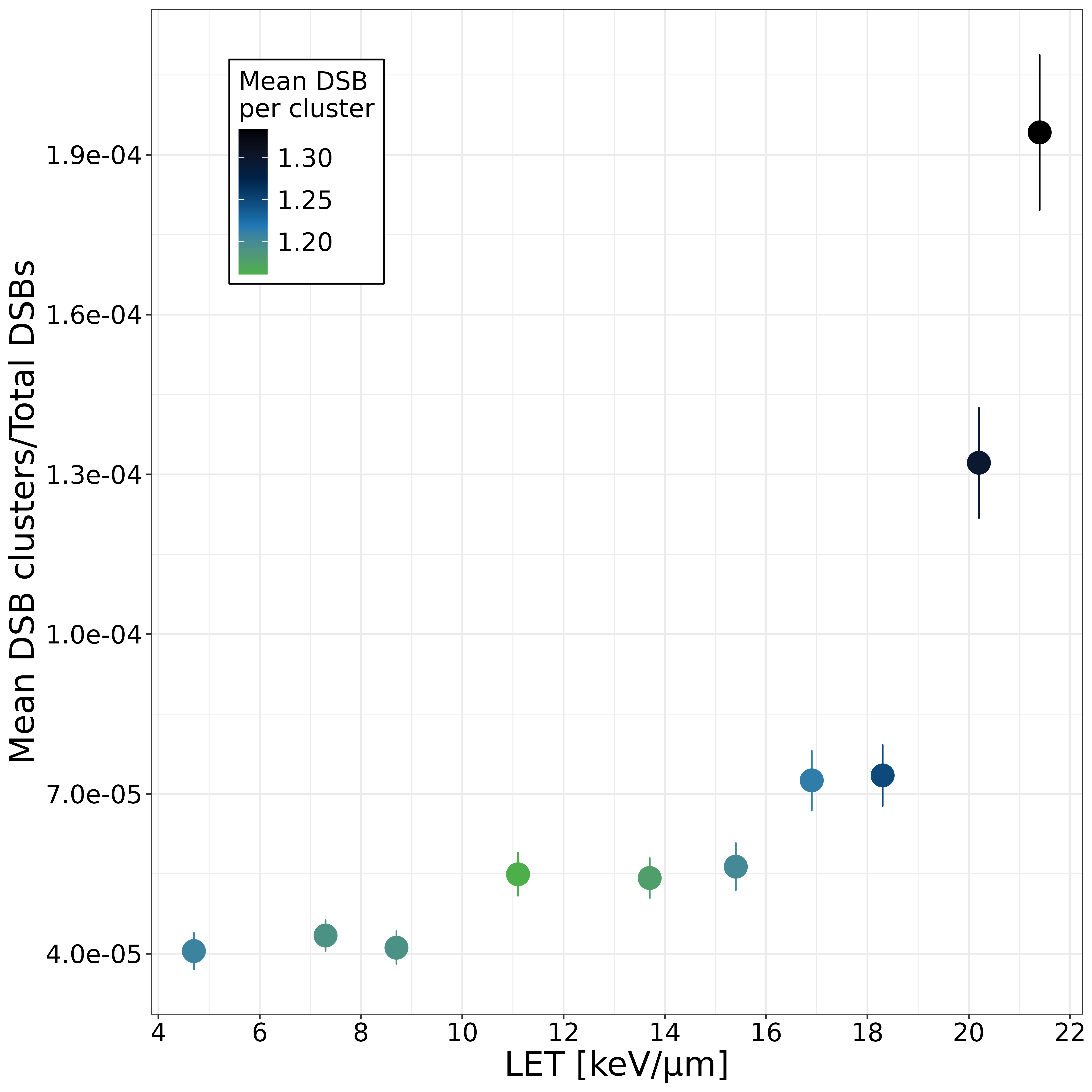}
    \caption{Cluster-to-total DSB ratio vs LET. The colormap reflects mean DSBs per cluster, i.e. mean cluster size.}
    \label{Fig. DSB_cluster}
\end{figure}

Figure \ref{Fig. DSB_cluster} illustrates the impact of nanodosimetry on radiation-induced biological effectiveness. Specifically, we report the ratio between the mean number of DSB clusters per cell and the total number of DSBs, plotted as a function of LET. The color scale encodes the average number of DSBs within each cluster, serving as an indicator of cluster size. An overall increasing trend is observed, indicating that both the number of clusters and the average cluster size grow with LET, reflecting the increased spatial correlation of DNA damage at higher ionization densities.

\begin{figure}[!h]
    \centering
    \includegraphics[width=\textwidth]{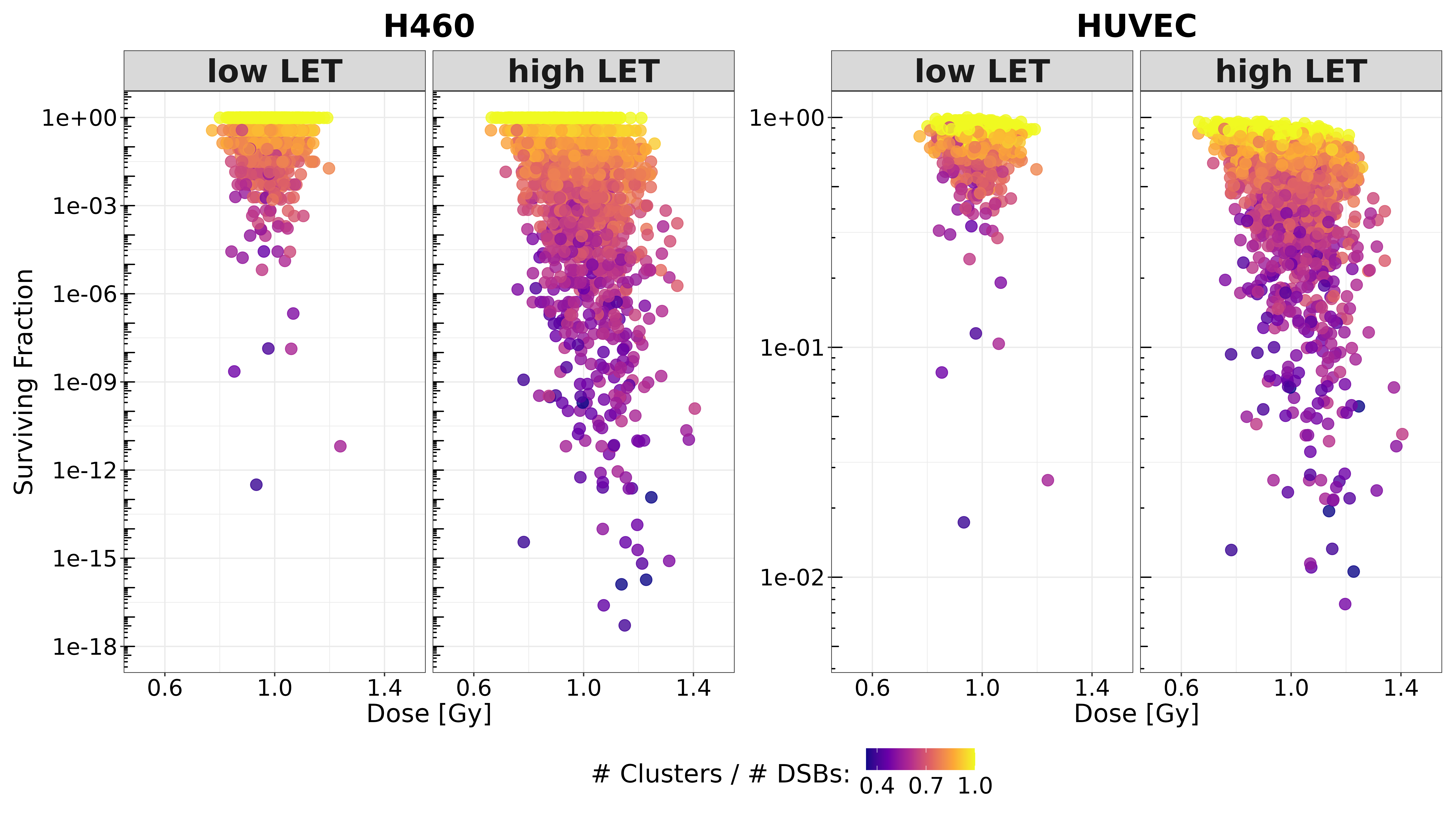}
    \caption{Surviving fraction plotted against absorbed dose centered at $1~Gy$ for individual simulations of proton-irradiated nuclei. Left: H460 cells; right: HUVEC cells. Irradiations are performed at LET$= 5~keV/\mu m$ (low LET), and $20~keV/\mu m$ (high LET). The colormap indicates the ratio of number of clusters to the total number of DSBs.}
    \label{Fig. SF_Dose_LET_analysis}
\end{figure}

To further explore how damage clustering influences biological outcomes, Figure \ref{Fig. SF_Dose_LET_analysis} presents the SF on a logarithmic scale as a function of dose for each simulated cell nucleus. Results are shown for two different cell lines, H460 (left panel) and HUVEC (right panel), irradiated with protons at two LET values: $5~keV/\mu m$ and $= 20~keV/\mu m$. In each panel, data are centered around $D_{\text{abs}} = 1~Gy$ to highlight the spread due to microdosimetric stochasticity. The colormap encodes a clustering index defined as the ratio of number of clusters to total DSBs within each nucleus. Lower values of this index correspond to more severe clustering (i.e., more DSBs per cluster). This analysis reveals a correlation between high clustering and reduced cell survival, particularly pronounced at higher LET values. This supports the central role of nanodosimetric features in shaping the biological response to ionizing radiation.

\subsection{\gsm Parameters}
\label{Subsubsection: parameters results}
Table \ref{tab:GSM2_parameters} reports the fitted \gsm parameters for the studied cell lines, along with the LQ model parameters for X-ray reference radiation used in the calculation of RBE$_{10}$ in proton irradiation experiments and simulations.
For the H460 cell line, \gsm parameters were obtained by fitting the model exclusively to the surviving fraction (SF) data at LET = 11.1keV/$\mu$m, as described in Section \ref{1_Materials_Methods}. No further adjustments were made to fit SF curves at other LET values, allowing the model to predict biological outcomes under varying irradiation conditions with fixed parameters.

\begin{table}[h]
    \centering
\begin{adjustbox}{max width=\textwidth}
\begin{tabular}{|c|c|c|c||c|c|}
  \hline
  \multirow{2}{*}{\textbf{Cell line}}
  & \multicolumn{3}{c||}{\textbf{\gsm parameters}}
  & \multicolumn{2}{c|}{\textbf{Reference radiation}} \\
  \cline{2-6}
  & $a$ [1/hr] & $b$ [1/hr] & $r$ [1/hr]&
  X-rays $\alpha [1/Gy]$ & X-rays $\beta [1/Gy^2]$ \\
  \hline
  \textbf{H460} & $6.45 \times 10^{-3}$ & $2.90$ & $1.74$ & $0.29$ & $0.083$ \\
  \hline
  \textbf{Endothelial} & $7.21 \times 10^{-2}$ & $2.83 \times 10^{-1}$ & $4.84$ & $0.41$ & $0.076$ \\ 
  \hline
  
\end{tabular}
\end{adjustbox}
\caption{Fitted \gsm parameters for the studied cell lines. LQ model parameters for X-ray reference radiation are used to compute RBE$_{10}$ for proton irradiation of H460 cells, following the protocol in \cite{guan2015spatial}.}
\label{tab:GSM2_parameters}
\end{table}

\section{Discussion}
\label{3_Discussion}
\counterwithin{figure}{section}
\setcounter{figure}{0}

\subsection{Nanodosimetry and Microdosimetry in a Radiobiological Model}
\label{Subsubsection: discussion model}
This work introduces \mtgsm, a novel framework that integrates the \gsm with the MINAS-TIRITH tool to predict radiation-induced biological effects. The model provides a fully stochastic, mechanistically grounded approach for simulating cell survival and RBE by explicitly coupling microdosimetric energy deposition with nanodosimetric DNA damage patterns.
\\
\\ 
A key innovation lies in quantifying how nanodosimetric events, in particular DSB formation and clustering,  govern radiobiological effectiveness. Figure~\ref{Fig. DSB_cluster} shows a strong correlation between the number of DSB clusters and RBE$_{10}$, as further supported by the trends in Figure~\ref{Fig. RBE_H_H460}. By incorporating spatial information at the scale of individual chromosomes, our approach enables biologically meaningful constraints: only DSBs occurring on the same chromosome can cluster. This spatial fidelity, enabled by MINAS-TIRITH, improves mechanistic modeling of repair dynamics and cell fate, advancing beyond previous approaches lacking this resolution.
\\
\\ 
The stochastic nature of both micro- and nanodosimetry is reflected in the variability of $z_n$ values (microdosimetric specific energy deposition) and the spatial clustering  of DSBs (nanodosimetric damage). As shown in Figure \ref{Fig. SF_Dose_LET_analysis}, lower LET radiation yields more homogeneous dose distributions, while higher LET introduces increased heterogeneity. At the nanoscale, high LET also leads to denser, more complex DNA damage, reducing survival probabilities, as highlighted by the clustering patterns and purple points associated with lower SF values. MINAS-TIRITH's further enables the distinction between direct, indirect and hybrid DSBs (Figure~ \ref{Fig. DSB_dir_indir}). As LET increases, radical recombination limits chemically induced damage, consistent with an increase in hybrid and direct DSBs. This supports the observation that at higher LET physical interactions dominate damage induction.
\\
\\
The application of DBSCAN within biologically realistic constraints offers a novel interpretation of the subdomain concept in theoretical radiobiological models. Unlike traditional models that rely on fixed-size artificial domains (e.g., in MKM), our model derives DSB clusters directly from spatial damage coordinates. This enhances biological realism and allows better prediction of cell survival across a broad range of radiation qualities. Importantly, the resulting RBE$_{10}$ values for protons often exceed the standard clinical value of 1.1, aligning more closely with experimental findings and reinforcing the role of nanodosimetry in biological effectiveness estimation.

\subsection{Model Applicability Across Radiation Types}

A major strength of \mtgsm is its predictive capability for particle and photon radiation. Unlike the MKM and the Local Effect Model (LEM) \cite{friedrich2012calculation, pfuhl2022comprehensive}, the only radiobiological model used in clinics, which requires externally calibrated radiosensitivity parameters for the reference radiation, \mtgsm predicts the biological effects of photons and charged particles from first principles. This is shown by the accurate reproduction of experimental SF data for photons (Figure \ref{Fig. SF_Xrays}) and improved residuals compared to the TLK model (Figure \ref{Fig. Res_Xrays}).
\\
\\ 
For proton irradiation, \mtgsm replicates experimental RBE trends. As shown in Section \ref{2_Results}, predicted RBE$_{10}$ values can exceed 3 at LET values around $20~keV/\mu m$, highlighting the substantial underestimation implied by the standard clinical assumption of RBE =$1.1$. This underestimation is particularly concerning near the Bragg peak’s distal edge, where normal tissues may receive high-LET exposure. Our findings support a shift toward spatially resolved, biologically informed RBE modeling in treatment planning, as also discussed in recent literature \cite{paganetti2024nrg}. 
\\
\\
It is important to emphasize that the commonly used RBE value of 1.1 was not selected based on precise biological evidence, but rather as a conservative and pragmatic estimate \cite{paganetti2024nrg}. However, the substantial variability in proton RBE, both along the beam path and laterally, suggests that more accurate modeling of biological effectiveness could lead to improved tumor control and reduced normal tissue toxicity. Incorporating spatially resolved, mechanistically derived RBE values into treatment planning may represent a critical step toward more personalized and biologically optimized radiotherapy.
In this work, we highlight the importance of including all the levels of stochasticity inherently given by radiation-matter interaction to formulate a model that can predict the RBE in such a wide range. In this context, \cite{paganetti2024nrg} has also proposed that proton therapy treatment planning could benefit from considering DNA DSB induction directly, rather than relying solely on cell survival as the biological endpoint. The approach presented here aligns precisely with this vision: \mtgsm directly connects DSB induction patterns to cell fate, considering DNA damage endpoints and survival curves alone. Further, its computational efficiency ensures the feasibility of integration into clinical workflows.
\\
\\
While small discrepancies remain, in particular in distal Bragg peak regions where dose gradients are steep, the model offers a substantial improvement over earlier approaches \cite{bordieri2024validation}, especially for high-LET protons ($\sim 20~keV/\mu m$).

\subsection{Analysis of \mtgsm parameters}
\label{Subsubsection: discussion model}
The \gsm parameters (Table \ref{tab:GSM2_parameters}), fitted to SF data for $220~kV$ X-rays and $11.1~keV/\mu m$ protons, reflect the radiosensitivity and repair dynamics of different cell lines. Parameters $a$, $b$, and $r$ govern the balance between repair and cell death mechanisms. \gsm assumes these three parameters to be cell-line-specific, which relates to the different biological responses to radiobiological damage. 
\\
\\
A consistent feature across cell lines is that $b$ (second-order term) exceeds $a$ (first-order term), emphasizing the importance of damage clustering. Moreover, $b$ and $r$ are now comparable in magnitude—a shift from previous models based on the MKM formalism, where $b$ was often negligible. This is due to a significant change from previous domain definitions in both MKM \cite{hawkins2003microdosimetric} and \gsm \cite{bordieri2024validation}. This change results from a biologically grounded redefinition of domains: our model uses fewer, physically motivated DSB clusters instead of many small artificial domains. This increases the significance of damage interactions within clusters, reinforcing $b$’s role in shaping survival outcomes. Consequently, the richness and predictive capacity of MT-\gsm arise from the dynamic interplay between repair processes ($r$), the first-order parameter ($a$), the second-order parameter ($b$), together with the power and level of detail from the MINAS-TIRITH tool, and the damages clustering process operated through DBSCAN.
\\
\\
Regarding the repair rate specifically, its role is crucial, and we consider the value of $r$ to be a measure of cell-line radiosensitivity. Table \ref{tab:GSM2_parameters} displays the parameters in ascending order for $r$, therefore from the more radiosensitive cell line to the more radioresistant, and these results confirm what is typically observed in in vitro experiments \cite{deschavanne1996review}. Moreover, the repair rate values align with the typical time for repairing the complex DNA damage in human cells \cite{fowler1999repair, carabe2011repair, dexheimer2012dna}.

\subsection{\mtgsm Outlook and Future Directions}
\label{Subsubsection: discussion model}
This study establishes a direct link between microdosimetric energy deposition and nanodosimetric DNA damage patterns, with potential implications for advanced radiotherapy treatment design. The main limitation is the extent of the MINAS-TIRITH database beyond currently available particle energies. Ongoing work aims to expand its coverage to include i) protons up to $\sim300~MeV$, covering full clinical ranges; ii) helium ions, offering intermediate LET characteristics, and iii) carbon ions, already in progress through updated Geant4-DNA set of cross-sections implementations.
These developments will enhance the model’s applicability in heavy ion therapy and enable deeper exploration of LET-dependent radiobiological effects in tumors and surrounding healthy tissues. As the framework evolves, it holds promise for incorporation into treatment planning systems, potentially enabling real-time adaptive, biologically optimized radiotherapy.

\section{Conclusions}
\label{4_Conclusions}
\counterwithin{figure}{section}
\setcounter{figure}{0}

In this work, we introduced \mtgsm, a novel radiobiological model that integrates the \gsm framework with the MINAS-TIRITH tool, enabling a unified multiscale approach that tightly connects microdosimetry and nanodosimetry for the prediction of cell survival. The model was extensively validated against experimental data from \cite{patel2017optimization} and \cite{bronk2020mapping}, demonstrating its ability to reproduce SF curves, LQ parameters and RBE$_{10}$ values for H460 cells irradiated with protons, as well as survival probabilities for HUVEC cells exposed to X-rays. Validation results show very good agreement with experimental data across a wide range of LET values, from approximately $4~keV/\mu m$ up to over $20~keV/\mu m$.
\\
\\
Our model establishes a mechanistic link between the stochastic structure of radiation tracks and DNA damage induction at the nanoscale, particularly focusing on the spatial clustering of DSBs through the DBSCAN algorithm. By providing a physical interpretation of the nucleus subdomain concept, \mtgsm offers a meaningful alternative to traditional radiobiological models currently used in clinical settings.
\\
\\
The model’s flexibility and generality—applicable across different radiation sources and cell types—enable it to capture the full stochastic nature of radiation interactions. Notably, this leads to RBE predictions for high-LET protons significantly greater than the conventional clinical assumption of $1.1$. This underlines the importance of accounting for nanodosimetric detail to avoid underestimating biological effects, particularly in the context of normal tissue sparing during treatment planning.
\\
\\
Finally, the extensibility of the model paves the way for predictive applications beyond existing experimental datasets. Future developments, including the expansion of the MINAS-TIRITH databases to include higher-energy protons and heavier ions such as helium and carbon, will broaden the model's applicability. The integrated micro- and nanodosimetric framework presented here lays a strong foundation for the advancement of personalized and precision radiotherapy.

\section*{Conflict of Interest Statement}
\label{6_Conflict_of_Interest_Statement}
\counterwithin{figure}{section}
\setcounter{figure}{0}
The authors have no relevant conflicts of interest to disclose.


\bibliographystyle{elsarticle-num} 
\bibliography{bib}

\end{document}